\newcommand{\be}{\begin{equation}}\newcommand{\ee}{\end{equation}}
\newcommand{\bea}{\begin{eqnarray}}\newcommand{\eea}{\end{eqnarray}}
\newcommand{\pl}{\partial}
\newcommand{\nn}{\nonumber}\newcommand{\p}[1]{(\ref{#1})}
\documentstyle[12pt]{article}
\begin{document}

\thispagestyle{empty}
\begin{flushright}
JHU-TIPAC-94011\\
hepth@xxx/9406217\\
June 1994
\end{flushright}

\bigskip\bigskip\begin{center} {\bf
\Large{Matter couplings in partially broken extended supersymmetry} }
\end{center} \vskip 1.0truecm

\centerline{\bf J.  Bagger${}^1$ and A.  Galperin${}^{1,2}$}
\vskip5mm
\centerline{Department of Physics and Astronomy}
\centerline{Johns Hopkins University}
\centerline{Baltimore, MD 21218, USA}
\vskip5mm

\bigskip \nopagebreak \begin{abstract}
\noindent We use nonlinear realizations to describe the spontaneous breaking
of $N=2$ supersymmetry to $N=1$ in four dimensions.  We identify the
Goldstone multiplet with an $N=1$ chiral superfield, and show that chiral
$N=1$ matter is consistent with the partially broken $N=2$ supersymmetry.
We find that the chiral matter can be in any representation of the gauge
group; no mirror particles are required.  We present the Goldstone action
and the general couplings to $N=1$ matter to the first nontrivial order in
the scale of symmetry breaking.
\end{abstract}

\bigskip \nopagebreak \begin{flushleft} \rule{2
in}{0.03cm} \\ {\footnotesize \ ${}^1$ Supported in part by NSF grants
PHY90-96198 and GER93-54057, and by Texas National Research Laboratory
Commission under grant RGFY93-292.}
\\ {\footnotesize \ ${}^2$ World Laboratory
Fellowship.}
\end{flushleft}

\newpage\setcounter{page}1

\noindent{\bf 1.  Introduction.}
\vskip 5mm
\noindent
In four dimensions, supersymmetric theories can be classified by an
integer $N$ that counts the number of supersymmetries.  $N$ can
run from one to eight, but most phenomenological studies focus on
$N=1$.  This is because $N=1$ is the simplest supersymmetry, and the
only one that permits fermions to lie in complex representations of
symmetry groups \cite{witten}.  Therefore if supersymmetry is relevant
to physics at the weak scale, it is likely to be $N=1$, broken to
$N=0$.

One cannot help but wonder whether this $N=1$ supersymmetry might be
a remnant of some $N>1$ extended supersymmetry, broken to $N=1$ at a
higher scale.  There is a general argument which implies that extended
supersymmetry cannot be spontaneously broken to $N=1$ in four dimensions.
The argument runs as follows \cite{witten}: Suppose that there are two
supersymmetries, one broken and one unbroken.  Since one supersymmetry is
preserved, its supercharge must annihilate the vacuum.  Because of the
supersymmetry algebra, the Hamiltonian must also annihilate the vacuum.
This implies that the other supercharge must annihilate the vacuum, so
the second supersymmetry cannot be broken.

Hughes, Liu and Polchinski \cite{pol,hughes} showed
how to evade this argument.  They considered $N=2$ supersymmetry,
spontaneously broken to $N=1$, and identified the Goldstone
multiplet with a four-dimensional membrane propagating in
six-dimensional superspace.  They found its invariant action, and
demonstrated that this system realizes the partial breaking of extended
supersymmetry.  This was a remarkable result, but one that was difficult
to use because of the membrane approach.  In particular, it was not clear
how to define chiral $N=1$ matter on the membrane, nor whether an
invariant matter action could be constructed.

These shortcomings motivated us to reconsider the partial breaking of
$N=2$ supersymmetry.  We do not use a membrane, but instead we work in
four-dimensional $N=1$ superspace and use the techniques of nonlinear
realizations to realize the second supersymmetry.  (For early work along
these lines, see \cite{BW}.) This approach keeps the unbroken
$N=1$ supersymmetry manifest, and allows $N=2$ invariants
to be constructed with the help of the Goldstone multiplet.
It permits us to examine the couplings of the Goldstone multiplet to
$N=1$ supersymmetric matter, and study any restrictions on $N=1$
matter that come from the second supersymmetry.

In this paper we will show that the Goldstone multiplet of partially
broken $N=2$ supersymmetry is described
by an $N=1$ chiral superfield, and that the chiral $N=1$ representation is
preserved by the second supersymmetry.  We will prove that any $N=1$
matter can be consistently coupled to the Goldstone multiplet.  We
will present the general Goldstone-matter coupling to the first nontrivial
order in the supersymmetry breaking parameter.  The underlying complex
geometry and full nonlinear structure will be discussed elsewhere
\cite{BG}.

By construction, the Goldstone-matter coupling exhibits $N=2$
supersymmetry, spontaneously broken to $N=1$.  We will see that when the
$N=1$ superpotential is $R$-symmetric, the theory also exhibits an
$SO(5,1)$ symmetry, spontaneously broken to
$SO(3,1)\times SO(2)$.  This extra
symmetry provides a hint about the origin of $R$ symmetry and the partial
breaking of extended supersymmetry.

\vskip5mm

\noindent{\bf 2.  The Goldstone multiplet and nonlinear realizations.}
\vskip5mm

\noindent The $N=2$ supersymmetry algebra can be written in the following
form,
\bea
\label{algebra}
\{Q_\alpha, \bar Q_{\dot\alpha}\}\! &=&\! 2\sigma^a_{\alpha\dot\alpha}P_a, \;\;
\{S_\alpha, \bar S_{\dot\alpha}\}=2\sigma^a_{\alpha\dot\alpha}P_a, \nn\\
\{Q_\alpha, S_\beta\}\! &=&\! 2\epsilon_{\alpha\beta}Z, \;\;
\{\bar Q_{\dot\alpha}, \bar S_{\dot\beta}\}=
2\epsilon_{\dot\alpha\dot\beta}\bar Z,
\eea
where $Q_\alpha$ and $S_\alpha$ are the supersymmetry generators, $P_a$
the four-dimensional momentum operator, and $Z$ is a complex central
charge, which can be viewed as translation generator along two additional
spacelike directions, $Z=P_4-i P_5$.  In what follows, we take $Q_\alpha$
to be the unbroken $N=1$ supersymmetry generator and $S_\alpha$ to be
its broken counterpart.

In general, spontaneous supersymmetry breaking gives rise to a massless
spin-1/2 Goldstone field $\psi_\alpha(x)$ \cite{VA}.  When $N=2$
supersymmetry is broken to $N=1$, the Goldstone fermion is part of a
massless $N=1$ supersymmetry multiplet.  There are two such multiplets that
contain spin 1/2: the chiral multiplet $(1/2, 0)$, and the vector
multiplet $(1, 1/2)$.  The supersymmetry anticommutator $\{Q_\alpha,
S_\beta\}=2\epsilon_{\alpha\beta}Z$ motivates us to choose the chiral
multiplet, whose complex spin-0 field $\phi(x)$ can be interpreted as a
Goldstone boson for the central charge generator of the supersymmetry
algebra.

Note that the {\it geometrical} dimension of a Goldstone field is opposite
to that of the corresponding broken generator.  Since $[Q]=[S]=1/2, \;
[Z]=1$, we find $[\psi]=-1/2, \; [\phi]=-1$, which is in accord with the
$N=1$ transformation law $\delta\phi =\epsilon^\alpha\psi_\alpha$.
In what follows, we will show that
it is consistent to take the chiral $N=1$ multiplet
$(\phi,\psi_\alpha)$ to be the Goldstone multiplet of the partially broken
$N=2$ supersymmetry.  This result was obtained in \cite{pol,hughes} using
different arguments.

In addition to the physical fields $(\phi,\psi_\alpha)$, the off-shell
Goldstone multiplet contains a complex auxiliary field $\bar\xi(x)$.
The $N=1$ transformation law $\delta\psi_\alpha=\epsilon_\alpha\bar\xi+\ldots$
implies that the geometrical dimension of $\bar\xi$ is 0.  As we will see, this
field can be interpreted as a Goldstone boson parametrizing the coset
$SU(2)/U(1)$,
where $SU(2)$ is an automorphism group of $N=2$ supersymmetry.  ($Q_\alpha$ and
$S_\alpha$ form an $SU(2)$ doublet, while $P_a$ and $Z$ are $SU(2)$ singlets).
Since the auxiliary field equation of motion is $\bar\xi=0$, the Goldstone
action explicitly breaks $SU(2)$ to $U(1)$.  Nevertheless, as we will
see later, it is useful to keep track of this $SU(2)$ group.

The formalism of nonlinear realizations provides a systematic way to study
the properties of Goldstone fields.  This formalism was first developed for
internal symmetries \cite{CCWZ} and later generalized to space-time symmetries
\cite{space}.  The procedure is as follows: Let $G$ be the full symmetry
group, and $H$ the unbroken subgroup.  The generators of $G$ can be divided
into three sets: space-time generators $\Gamma_{A}$, spontaneously broken
internal generators $\Gamma_r$, and the unbroken generators $\Gamma_i$ of
the subgroup $H$.  The $\Gamma_i$ form an ordinary Lie algebra, $[\Gamma_i,
\Gamma_j]={\cal C}_{ij}{}^k \Gamma_k$, while the generators $\Gamma_{A}$ and
$\Gamma_r$ span representations of $H$, $[\Gamma_i,\Gamma_{A}] = {\cal
C}_{i{A}}{}^{B}\Gamma_{B}, \;\;[\Gamma_i, \Gamma_r]={\cal C}_{ir}{}^s
\Gamma_s.$

For any group $G$ and subgroup $H$, we choose to parametrize the coset $G/H$
as follows,
\be
\label{coset}
\Omega=\exp (iX^{A}\Gamma_{A})
\exp (i\xi^r\Gamma_r),
\ee
where the $X^{A}$ are spacetime coordinates and $\xi^r=\xi^r(X)$ are the
Goldstone fields that correspond to the broken generators $\Gamma_r$.  Then
the group $G$ can be realized on the space-time coordinates and Goldstone
fields in the following way,
\be
\label{Greal}
g\Omega=\Omega' h,
\ee
where $g \in G$, $\Omega'=\exp (iX'^{A}\Gamma_{A}) \exp
(i\xi'^r(X')\Gamma_r) $.
In this expression, $h=\exp (i\lambda^i(g, X, \xi)\Gamma_i)$ is an element of
$H$ that is chosen to restore the form of $\Omega$.  With these rules, the
coordinates and Goldstone fields transform linearly under $H$.

Given a field $\chi(X)$ that transforms linearly under $H$, one can
define a realization of $G$ on $\chi(X)$ using the element $h$,
\be
\chi'(X')=\exp (i\lambda^i(g, X, \xi)\Gamma_i)\;\chi(X).
\ee
The generators $\Gamma_i$ are in the appropriate representation of $H$.
Note that this is a covariant transformation law for all elements $g \in G$.

To construct an invariant action, it is useful to define connection
and vielbein forms with the help of the Cartan one-form, $\Omega^{-1}
d\Omega$.  The Cartan form can be expanded with respect to the $G$ generators
\be
\Omega^{-1}d\Omega=i(\omega^{A}\Gamma_{A} +
\omega^r\Gamma_r + \omega^i\Gamma_i)
\ee
to give the forms $\omega^{A}, \omega^r$ and $\omega^i$.
{}From
\be
(\Omega^{-1}d\Omega)'=h(\Omega^{-1}d\Omega)h^{-1} +h d h^{-1},
\ee
we see that $\omega^{A}$ and $\omega^r$ transform homogeneously under $G$,
while $\omega^i$ transforms by a shift.

The vielbein $E_{M}{}^{A}$ is obtained by expanding the space-time form
$\omega^A$ with
respect to the coordinate differentials $dX^M$,
\be
\label{vielbein}
\omega^{A}=dX^{M}E_{M}{}^{A}.
\ee
The covariant derivatives of the Goldstone fields are found by expanding
the Goldstone forms $\omega^r$ with respect to $\omega^{A}$
\be
\omega^r= \omega^{A}\tilde{\cal D}_{A}\xi^r .
\ee
Finally, the connection one-form $\omega^i$ can be used to construct covariant
derivatives of the fields $\chi$,
\be
{\cal D}\chi =\omega^{A}{\cal D}_{A}\chi = (d+i\omega^i\Gamma_i )\chi.
\ee
These are the building blocks that can be used to construct actions invariant
under $G$.

\vskip 5mm

\noindent{\bf 3.  Goldstone constraints and action.}
\vskip5mm
\noindent To apply the above formalism to partially broken $N=2$ supersymmetry,
we
need to specify the groups $G$ and $H$.  The group $G$ must contain the
supersymmetry transformations \p{algebra}, but it can also
include various automorphisms of the supersymmetry algebra.
The maximal automorphism group is $SO(5,1)\times SU(2)$,
where $SO(5,1)$ is the $D=6$ Lorentz group.  (Under $SO(5,1)$, the
generators $P_a$ and $Z$ form a $D=6$ vector, while the supercharges
form a $D=6$ Majorana-Weyl spinor).  We denote the
maximal group $G$ as $G_{max}$.
The group $H\subset G$ should act linearly on the Goldstone fields.
The maximal linear group acting on the Goldstone multiplet
$(\phi, \psi_\alpha, \bar\xi)$ is a subgroup of $G_{max}$
\be
H_{max}=SO(3,1)\times SO(2)\times U(1),
\ee
where $SO(3,1)\times SO(2) \subset SO(5,1)$, $U(1)\subset SU(2)$, and
$SO(3,1)$ is the $D=4$ Lorentz group.

In what follows, we take $G/H=G_{max}/H_{max}$.  As
we will see, this choice is consistent with $N=1$ chirality and K\"ahler
invariance.  (In section 5, we will consider other cosets $G/H$.  We will
see that chirality and K\"ahler invariance imply $G/H=G_{max}/H_{max}$.)

A parametrization of the coset $G_{max}/H_{max}$
involves the $N=1$ superspace coordinates $X^{A}=(x^a,
\theta^\alpha, \bar\theta_{\dot\alpha})$, as well as the Goldstone
superfields $\Phi(x,\theta,\bar\theta), \ldots, \bar\Xi
(x,\theta,\bar\theta)$:
\bea
\Omega&=&\exp i(x^aP_a+\theta^\alpha Q_\alpha
+\bar\theta_{\dot\alpha}\bar Q^{\dot\alpha}) \exp i(\Phi Z+\bar\Phi\bar Z+
\Psi^\alpha S_\alpha+\bar\Psi_{\dot\alpha}\bar S^{\dot\alpha}) \nn\\
&&\times\exp i(\Lambda^aK_a+ \bar\Lambda^a\bar K_a +\Xi T
+\bar\Xi\bar T).
\eea
Here $\Lambda^a, \bar\Lambda^a$ are the Goldstone superfields associated
with the generators $K_a, \bar K_a$ of\break $SO(5,1)/SO(1,3)\times SO(2)$,
\be
[K_a, \bar K_{b}]=-2iL_{ab} -2\eta_{ab}M, \;\; [K_a, K_b]=
[\bar K_a, \bar K_b]=0,
\ee
where $L_{ab}$ and $M$ generate $SO(1,3)\times SO(2)$.  Similarly, $\Xi,
\bar\Xi$ are the Goldstone superfields for the broken generators
$T, \bar T$ of $SU(2)/U(1)$,
\be
[T, \bar T]=2T_0, \;[T_0, T]=T, \;[T_0,\bar T]=- \bar T.
\ee

The symmetry transformations of the Goldstone superfields
follow from \p{Greal} and the relations,
\bea\label{D=6 supersymmetry}
&&[K_a, P_b]=i\eta_{ab}Z, \;\; [K_a, Z]=0,\;\; [K_a, \bar Z]=2i P_a; \nn\\
&&[K_a, \bar Q^{\dot\alpha}]=i\bar\sigma_a^{\dot\alpha\alpha}S_\alpha, \;\;
[K_a, \bar S^{\dot\alpha}]=-i\bar\sigma_a^{\dot\alpha\alpha}Q_\alpha, \\
&&[K_a, Q_\alpha]=[K_a, S_\alpha]=0;\nn\\
&&[T, Q_\alpha]=0, \;\; [T, S_\alpha]= Q_\alpha, \;\;
[T, \bar Q^{\dot\alpha}]=- \bar S^{\dot\alpha},\;\;
[T, \bar S^{\dot\alpha}]=0 .\nn
\eea
In particular, to lowest order in the Goldstone fields, they are:
\vskip5mm
\noindent
Second supersymmetry ($g=\exp i(\eta S+\bar\eta\bar S)$)
\be\label{S}
\delta\Phi=2i \eta\theta, \; \delta\Psi_\alpha=\eta_\alpha, \;\; \delta
\Lambda=\delta\Xi=0.
\ee
\noindent
$K, \bar K$ transformations ($g=\exp i(rK+\bar r\bar K)$)
\bea\label{K}
\delta\Phi&=&-r_a(x^a-i\theta\sigma^a\bar\theta)+\ldots,\;\;
\delta\Psi^\alpha=-r_a(\bar\theta\bar\sigma^a)^\alpha, \nn\\
\delta\Xi&=&0, \;\;
\delta \Lambda^a=r^a +\ldots.
\eea
\noindent
$T, \bar T$ transformations ($g=\exp i(\beta T+\bar\beta\bar T)$)
\bea\label{T}
\delta\Phi=-\bar\beta\theta\theta+\ldots, \;\;
\delta\Psi^\alpha=i\bar\beta\theta^\alpha\nn\\
\delta \Lambda^a=0, \;\; \delta\xi=\beta +\ldots.
\eea

The $N=1$ superfields $\Phi(x, \theta, \bar\theta), \ldots,
\bar\Xi(x, \theta, \bar\theta)$ contain many more components than the
physical Goldstone multiplet $\phi(x), \psi^\alpha(x), \bar\xi(x)$.
Therefore the superfields must be properly constrained.  To this end,
we note that no $G_{max}$ tensors of dimension $-1, -1/2$ or 0 can be
built from the physical fields.  This motivates us to impose the following
constraints \cite{similar}:
\bea
&&{\tilde{\bar{\cal D}}}_{\dot\alpha}\Phi=0,\;\;
{\tilde{\cal D}}_{\alpha}\Phi=0,\;\;
{\tilde{\cal D}}_{a}\Phi=0\;\; \label{constr1} \\
&&{\tilde{\cal D}}_{\alpha}\Psi^\beta=0,\;\;
{\tilde{\bar{\cal D}}}_{\dot\alpha}\Psi^\beta=0.  \label{constr2}
\eea
Except for the trace part ${\tilde{\cal D}}^{\alpha}
\Psi_\alpha=0$, the constraints \p{constr2} are necessary to ensure the
consistency of \p{constr1}.

The constraints ${\tilde{\cal D}}_{\alpha}\Phi=0,\;
{\tilde{\cal D}}_{a}\Phi=0\; {\tilde{\cal D}}^{\alpha}\Psi_\alpha=0$
allow us to express the Goldstone superfields $\Psi^\alpha, \Lambda^a$ and
$\bar\Xi$ in terms of a single superfield $\Phi$ \cite{Ivanov}.
To lowest order, we find
\be
\Psi^\alpha=-{i\over 2}D^\alpha\Phi +\ldots; \;\;
\Lambda_a=-\pl_a\Phi +\ldots; \;\; \bar\Xi={1\over 4}D^2\Phi +\ldots.
\ee
The constraint ${\tilde{\bar{\cal D}}}_{\dot\alpha}\Phi=0$ reduces $\Phi$
to an $N=1$ chiral superfield (see \p{goldsolution}).  The constraints
\p{constr1}, \p{constr2} are
consistent with the transformations \p{S} -- \p{T}.

To second order, the perturbative solution to the Goldstone constraints
\p{constr1}, \p{constr2} is given by
\be
\label{goldsolution}
\Phi=\varphi -{i\over 4}(D\varphi\sigma^a\bar D\bar\varphi) \pl_a\varphi
-(\pl_a\varphi)^2\bar\varphi +O(\varphi^5),
\label{goldst}
\ee
where $\varphi=\varphi(x-i\theta\sigma\bar\theta, \theta)$ is an ordinary
$N=1$ chiral superfield, $\bar D_{\dot\alpha}\varphi=0$, with components
$\phi(x), \psi^\alpha(x), \bar\xi(x).$

The Goldstone action is uniquely determined by the
requirements of $Q$ and $S$ supersymmetries.  To this order, it is
\be
\label{goldaction1}
S_{g}={1\over a^2}\int d^4x d^2\theta d^2\bar\theta \;
[\varphi\bar\varphi
-{1\over 2}(\pl_a\varphi)^2\bar\varphi^2
-{1\over 2}(\pl_a\bar\varphi)^2\varphi^2-{1\over 16}D^\alpha\varphi
D_\alpha\varphi\bar D_{\dot\alpha}\bar\varphi\bar D^{\dot\alpha}\bar\varphi
+O(\varphi^6)].
\ee
It can be rewritten in terms of the original Goldstone superfields as follows,
\be
\label{goldaction}
S_{g}={1\over a^2}\int d^4x d^2\theta d^2\bar\theta \;
E \; [\Phi\bar\Phi
+{1\over 2}\Lambda^2\bar\Phi^2 +{1\over 2}\bar\Lambda^2\Phi^2 +
\Psi^2\bar\Psi^2 + O(\Phi^6)].
\ee
In this expression, $E\!=$Ber($E_{M}{}^{A}$) is the superdeterminant of
the vielbein, and $a$ is a constant of dimension 2 that corresponds to the
scale of the second supersymmetry breaking.  The action \p{goldaction1},
\p{goldaction} is invariant under $SO(5,1)$, but explicitly breaks $SU(2)$
down to $U(1)$.

\vskip5mm

\noindent{4.  \bf Matter constraints and invariant action.}
\vskip5mm
\noindent The above constraints imply that the chirality condition
\be
\label{chiralcond}
\bar{\cal D}_{\dot\alpha}\chi=0
\ee
is consistent for a ``matter'' superfield $\chi$ in an arbitrary
representation of $H_{max}$,
\be
\label{consistency}
\{\bar{\cal D}_{\dot\alpha}, \bar{\cal D}_{\dot\beta}\}\chi=0.
\ee

The solution to the chiral matter constraint \p{chiralcond} for an
$H_{max}$ singlet is given in terms of an arbitrary holomorphic function
\be
\chi=\chi(x_{L}, \theta_{L}),
\ee
where
\bea\label{holomorph}
x^a_{L}&=&x^a-i\theta\sigma^a\bar\theta -i\Psi\sigma^a\bar\Psi
+2i\Lambda_b\bar\theta\bar\sigma^a\sigma^b\bar\Psi+
2\Lambda_a\bar\Phi +O(\Phi^4),\nn\\
\theta_{L}^\alpha&=&\theta^\alpha-\Lambda^a(\bar\Psi\bar\sigma)^\alpha
+O(\Phi^4).
\eea
This set of holomorphic coordinates is closed under $G_{max}$
\cite{BG}; it generalizes the ordinary chiral $N=1$ superspace in the
presence of the Goldstone superfield.

The $G_{max}$-invariant kinetic term for $\chi$ is given by
\be
\label{kin}
S_{\rm kin}=\int d^4xd^2\theta d^2\bar\theta
\;E\; K(\bar\chi_i, \chi^j)
\ee
where $K$ is the K\"ahler potential.  The action \p{kin} is invariant under
K\"ahler transformations $K \rightarrow K +F(\chi)
+\bar F(\bar\chi)$.  This fact follows from a remarkable property of
the Goldstone supervolume: an invariant integral of a (covariantly) chiral
superfield vanishes identically,
\be
\label{vanish}
\int d^4x d^2\theta d^2\bar\theta \;E\; \chi(x_{L},
\theta_{L})=0.
\ee
This can be proven by passing from the real basis $(x, \theta, \bar\theta)$
to a holomorphic basis $(x_{L}, \theta_{L},\bar\theta)$.  In this basis the
superdeterminant is itself holomorphic, that is, it does not depend on
$\bar\theta$:
\bea
\label{chiralmeasure}
E_{L}&=& E\times {\rm Ber}{ \pl(x, \theta, \bar\theta)\over
\pl(x_{L}, \theta_{L},
\bar\theta)}\nn\\
&=&1-{1\over 8}\bar D^2(\bar\Phi D^2\Phi) + O(\Phi^4).
\eea
Because of this property, the superdeterminant $E_{L}$ can be
used as a density for the superpotential term
\be
\label{superpot}
S_{\rm superpot}=\int d^4x_{L} d^2\theta_{L}
\;E_{L}\; P(\chi^i).
\ee
The action \p{goldaction}, \p{kin} and \p{superpot} is invariant
under $N=2$ supersymmetry for any K\"ahler potential $K$ and
superpotential $P$.

If the matter action \p{kin}, \p{superpot} is invariant under a rigid
internal symmetry,
\be\label{internal}
\chi \rightarrow e^{i\lambda}\chi, \;\; \bar\chi
\rightarrow \bar\chi e^{-i\lambda},
\ee
it can be gauged by introducing an $N=1$ Yang-Mills superfield
$V(x, \theta, \bar\theta)$ that takes its value in the algebra of the
internal symmetry group.  Under a gauge transformation, $V$ transforms
as follows,
\be\label{ymtransf}
e^V \rightarrow e^{i\lambda}e^V e^{-i\bar\lambda}.
\ee
Here $\lambda=\lambda(x_{L}, \theta_{L})$ is an arbitrary
{\it holomorphic} gauge parameter, and $\bar\lambda$
is antiholomorphic.  The gauged $N=2$ supersymmetric action is obtained
by replacing
\p{kin} by
\be\label{kingauged}
S_{\rm kin}=\int d^4xd^2\theta d^2\bar\theta
\;E\; K(\bar\chi_i e^{-V}, \chi^j).
\ee
The Yang-Mills action coupled to the Goldstone superfield is given
by
\be\label{ymaction}
S_{\rm YM}=\int d^4x_{L} d^2\theta_{L}
\;E_{L}\;{\rm Tr}(W^{\alpha}
W_{\alpha}) + c.c
\ee
where $W_{\alpha}=i\bar{\cal D}^2 (e^V {\cal D}_\alpha e^{-V})$ is the
Yang-Mills field strength.  The actions \p{kingauged} and
\p{ymaction} are invariant under the full group
$G_{max}$.

It is interesting to note that if $K$ and $P$ are such that the
corresponding $N=1$ theory is $R$-invariant
(that is, invariant under rigid transformations $\theta\rightarrow
e^{i\alpha}\theta,
\;\chi^i \rightarrow e^{iq_i\alpha}\chi^i$), then the resulting $N=2$
theory can be made invariant under the full six-dimensional Lorentz
group $SO(5,1)$.  To see this, let us first consider the superpotential.
The chiral measure is invariant under both supersymmetries and
the $D=4$ Lorentz group, but
transforms as $(-1,-1)$ under the induced
$SO(2) \times U(1) \subset SO(5,1) \times SU(2)$
\be\label{chirtransf}
d^4x_{L} d^2\theta_{L}\rightarrow e^{-2i\alpha}(d^4x_{L} d^2\theta_{L}).
\ee
Here $\alpha=i\bar r\Lambda-i\bar\beta\Xi
+O(\Phi^3)$ is the {\it holomorphic} parameter
of the induced $SO(2) \times U(1)$.
If the superpotential carries $R$-charge 2, we
can identify the induced $SO(2)$ and $U(1)$ with $U(1)_R$ on
the matter fields.  In this way the superpotential term can be
rendered invariant under the full group $G_{max}$.

The kinetic term is a little more subtle.  Its measure is invariant under
$SO(2) \times U(1)$, so it is also invariant under the full group
$G_{max}$.  Since the parameter $\alpha$ is field-dependent,
$\bar\alpha\neq\alpha$, and the action \p{kingauged} is not invariant
under $SO(5,1) \times SU(2)$ unless there is an analog $U$ of an abelian
gauge superfield to compensate the difference,
\be\label{analog}
U \rightarrow U +i(\alpha-\bar\alpha).
\ee
Then the $G_{max}$ invariant kinetic term is
\be
S_{kin}=\int d^4x d^2\theta d^2\bar\theta \;E\; K(\bar\chi_i e^{-V-
q_i U}, \chi^j).
\ee
Such an abelian gauge field is built from
the K\"ahler potentials $K_{SO(5,1)}, K_{SU(2)}$ for the cosets
$SO(5,1)/SO(1,3)\times SO(2)$ and $SU(2)/U(1)$, respectively
\be
U= K_{SO(5,1)} - K_{SU(2)}; \;\;\;
K_{SO(5,1)}=\Lambda\bar\Lambda +
O(\Phi^4), \;K_{SU(2)}=\Xi\bar\Xi+O(\Phi^4).
\ee

{}To sumarize, we have seen that any $N=1$ matter lagrangian can be made
$N=2$
supersymmetric with the help of the Goldstone multiplet.  If the $N=1$
theory is $R$-invariant, the symmetry can be enlarged to include the
full $D=6$ Lorentz group $SO(5,1)$.  In this case the matter
coupling is $SU(2)$ symmetric, although the $SU(2)$ symmetry
is explicitly broken down to $U(1)$ by the Goldstone action itself.

\vskip5mm

\noindent{\bf 5.  A comment on other cosets.}
\vskip5mm
\noindent The above construction of the Goldstone multiplet and its
matter couplings is based on the coset $G_{max}/H_{max}$.  Let us now
briefly discuss the other possible cosets $G/H$ related to $N=2$
supersymmetry.  Unlike $G_{max}/H_{max}$, these cosets give rise
to {\it dimensionless} tensors.  Indeed, for $G\subset G_{max}$, the algebra
of $G$ does not involve the $SO(5,1)/SO(3,1)\times SO(2)$ generators $K,
\bar K$ or the $SU(2)/U(1)$ generators $T, \bar T$ (or both).
This implies the
existence of at least one dimensionless tensor, either
$\bar{\cal D}^{\dot\alpha}\Psi^\alpha
\sim \bar\sigma_a^{\dot\alpha\alpha}\pl_a\Phi+\ldots$ (if $K, \bar K$ are
excluded), or ${\cal D}^\alpha\Psi_\alpha\sim D^2\Phi+\ldots$ (if $T, \bar
T$ are excluded), or both.  The dots denote terms of higher order
in the Goldstone fields.

As a result, the usual $G/H$ construction \cite{CCWZ} is ambiguous
because the dimensionless tensors can be used to modify
the covariant derivatives ${\cal D}_\alpha,
\bar{\cal D}_{\dot\alpha}$.  One can show \cite{BG} that the
requirements of covariantly chiral $N=1$ superfields and K\"ahler
invariance restrict the nonminimal terms in just such as way as to
effectively restore the coset with the maximal symmetry
$G_{max}/H_{max}$.

\vskip 5mm

\noindent{\bf 6.  Conclusions.}
\vskip 5mm
\noindent In this paper we reformulated partially broken $N=2$ supersymmetry
in a manifestly $N=1$ supersymmetric way.  We showed that any
$N=1$ matter can be made $N=2$ supersymmetric with the help of the
Goldstone multiplet.  In particular, we found that $N=1$ chirality is
preserved in the presence of the Goldstone superfield.  This implies that
the fermions of this effective theory can be chiral, and no mirror
fermions are necessary.

It is still not clear whether this effective theory discussed here
can arise from a theory with linearly realized $N=2$ supersymmetry.
If such a theory exists, the $SO(5,1)$ symmetry hints that the linear
theory should most probably be formulated in six dimensions.

An unusual feature of the theories with partially broken supersymmetry is
the existence of {\it two} conserved energy-momentum tensors \cite{pol,hughes}.
This follows from the fact that in a supersymmetric theory, an
energy-momentum tensor can be obtained by anticommuting the supersymmetry
current with itself
\be
\{\bar Q_{\dot\alpha}, J^Q_{m\alpha}(x)\}=
\int d^3 \vec{y} \{\bar J^Q_{0\dot\alpha}(\vec{y}), J^Q_{m\alpha}(x)\}=
2\sigma^n_{\alpha\dot\alpha}T_{mn}(x).
\ee

In the case of partially broken $N=2$ supersymmetry,
there are two conserved supersymmetry currents
\bea
J^{Q}_{m\alpha}&=&-{2i\over a^2}
(\sigma^a\bar\sigma_m\psi)_\alpha\pl_a\bar\phi
+\ldots\nn\\
J^{S}_{m\alpha}&=&{2i\over a^2}(\sigma_m\bar\psi)_\alpha +\ldots
\eea
that give rise to two symmetric conserved energy-momentum tensors
$T^{Q}_{mn}$ and $T^{S}_{mn}$.  The two energy-momentum tensors differ
by a constant term,
\be
T^{S}_{mn}=T^{Q}_{mn}+{1\over a^2}\eta_{mn}.
\ee

This constant term leads to a new phenomenon in a theory of
spontaneously broken symmetry: the algebras of the broken and unbroken
supersymmetries differ.  This is why the ``no-go'' arguments of
\cite{witten} can be avoided.  However, the presence of two
energy-momentum tensors is a problem when coupling
these theories to supergravity.  Work along these lines is
in progress.

\vskip5mm

We gratefully acknowledge discussions with V.I.  Ogievetsky, J.  Polchinski,
and E.  Witten.


\newpage
\begin{thebibliography}{99}
\bibitem{witten} E.  Witten, {\sl Nucl.  Phys.} {\bf B188} (1981) 513.
\bibitem{pol} J.  Hughes, J.  Liu and J.  Polchinski,
{\sl Phys.  Lett.} {\bf 180B} (1986) 370.
\bibitem{hughes} J.  Hughes and J.  Polchinski, {\sl Nucl.  Phys.}
{\bf B278} (1986) 147.
\bibitem{BW} J.  Bagger and J.  Wess,
{\sl Phys.  Lett.} {\bf 138B} (1984) 105.
\bibitem{BG} J.  Bagger and A.  Galperin, Johns Hopkins preprint
JHU-TIPAC-94012.
\bibitem{VA} D.V.  Volkov and V.P.  Akulov,
{\sl JETP Lett.} {\bf 16} (1972) 438.
\bibitem{CCWZ} C.  Callan, S.  Coleman, J.  Wess and B.  Zumino,
{\sl Phys.  Rev.} {\bf 177} (1969) 2247.
\bibitem{space} D.V.  Volkov, {\sl Sov.  J.  Particles and Nuclei} {\bf 4}
(1973) 3.\\
V.I.  Ogievetsky, Proceedings of X-th Winter School of Theoretical
Physics\break in Karpacz, vol. 1, p. 227 (Wroclaw, 1974).
\bibitem{similar} Similar constraints have been considered in the context of
two-dimensional supersymmetry \cite{hughes}.
\bibitem{Ivanov} This is an example of the so-called ``inverse Higgs effect.''
See E.A.  Ivanov and  V.I. Ogievetsky, {\sl Teor. Mat. Fiz.}
{\bf 25} (1975) 164.

\end{thebibliography}
\end{document}